\documentclass[12 pt]{article}
\usepackage{amsmath}
\usepackage{graphicx}
\usepackage{url}
\usepackage{amsfonts}
\usepackage{amssymb}
\DeclareMathAlphabet{\cyrm}{U}{UWCyr}{m}{n}
\DeclareSymbolFont{cyrm}{U}{UWCyr}{m}{n}
\DeclareSymbolFontAlphabet{\cyrm}{cyrm}
\DeclareMathSymbol{\Evo}{\cyrm}{cyrm}{"03}
\newtheorem{theorem}{Theorem}

\newtheorem{lemma}[theorem]{Lemma}

\newtheorem{proposition}[theorem]{Proposition}
\newtheorem{remark}[theorem]{Remark}

\newenvironment{proof}[1][Proof]{\noindent\textbf{#1.} }{\
\rule{0.5em}{0.5em}}
\begin{document}

\title{Higher symmetries of the elliptic Euler-Darboux equation}
\author{D. Catalano Ferraioli, G. Manno, F. Pugliese}

\maketitle
\begin{abstract}
We find a remarkable subalgebra of higher symmetries of the
elliptic Euler-Darboux equation. To this aim we map such equation
into its hyperbolic analogue already studied by Shemarulin. Taking
into consideration how symmetries and recursion operators
transform by this complex contact transformation, we explicitly
give the structure of this Lie algebra and prove that it is
finitely generated. Furthermore, higher symmetries depending on
jets up to second order are explicitly computed.
\end{abstract}

\noindent\textbf{Keywords}: jet spaces, Euler-Darboux equation,
complex transformation, higher symmetries, recursion operators.

\smallskip
\noindent\textbf{MSC 2000 Classification: 58A20, 70S10, 35Q05}

\section*{Introduction}

In this paper we find a remarkable subalgebra of higher symmetries (also known
as generalized symmetries, \cite{Olver01}) of the following linear elliptic
equation:
\begin{equation}
\mathcal{E}_{ED}=\left\{  \left(  x+y\right)  \left(  u_{xx}+u_{yy}\right)
+u_{x}+u_{y}=0\right\}  \label{eqnostra}%
\end{equation}
in the unknown function $u=u\left(  x,y\right)  $. This Euler-Darboux type
equation appears in two recent papers (\cite{VVS1}, \cite{VVS2}) devoted to
the study of Lorentzian Ricci flat 4-metrics admitting a bidimensional
nonabelian Lie algebra $\mathcal{G}$ of Killing vector fields with non null
orbits. Coordinates $\left(  x,y\right)  $ appearing in (\ref{eqnostra}) are a
conformal chart on the Riemannian surface $M$, quotient of the space-time with
respect to Killing foliation (see below for further details). Hence,
(\ref{eqnostra}) must be considered as (a local representation of) a second
order equation on the trivial bundle $M\times\mathbb{R}\rightarrow M$,
$\left(  x,y,u\right)  \rightarrow\left(  x,y\right)  $.

Let us briefly recall the physical origin of (\ref{eqnostra}). It has been
shown in \cite{VVS1} that, if the distribution orthogonal to the orbits is
completely integrable and $\mathcal{G}$ admits a null Killing vector field
$Y$, then the most general Ricci flat metric of this type takes, in a suitable
chart $\left(  x_{1},x_{2},p,q\right)  $, the form%
\begin{equation}
g=2f\left(  dx_{1}^{2}+dx_{2}^{2}\right)  +\mu\left[  \left(  h\left(
x_{1},x_{2}\right)  -2q\right)  dp^{2}+2dpdq\right]  \text{,}
\label{vilasi_spara_vino}%
\end{equation}
where: $\mu=A\,\Phi+B$ with $A,B\in\mathbb{R}$, $\Phi(x_{1},x_{2})$ is a non
constant harmonic function, $f=\pm\left(  \nabla\Phi\right)  ^{2}%
/\sqrt{\left|  \mu\right|  }$ and $h\left(  x_{1},x_{2}\right)  $ is a
solution of equation
\begin{equation}
\Delta h+\left(  \partial_{x_{1}}\ln\left|  \mu\right|  \right)
\partial_{x_{1}}h+\left(  \partial_{x_{2}}\ln\left|  \mu\right|  \right)
\partial_{x_{2}}h=0. \label{pinkus}%
\end{equation}
Space-time metrics (\ref{vilasi_spara_vino}) were already known since the
earlier works of Dautcourt, Ehlers, Kramer and Kundt (see \cite{Dautcourt},
\cite{Ehlers-Kundt}, \cite{Kramer}, \cite{Kundt} and also Chapter 24 of
\cite{Steph-et-al}) on Ricci flat 4-metrics with a null Killing vector field.

In fact, one can distinguish two cases according to whether $\mu$ is constant
or not. When $\mu$ is constant the above metric is a particular case of
pp-wave (see \cite{Steph-et-al}). On the contrary, if $\mu$ is not constant,
by taking $\mu$ and its harmonic conjugate $\tilde{\mu}$ as new coordinates
one can bring (\ref{pinkus}) to the more simple form
\begin{equation}
\mu\left(  h_{\mu\mu}+h_{\tilde{\mu}\tilde{\mu}\text{ }}\right)  +h_{\mu}=0
\label{pinkus_quasi_pallinus}%
\end{equation}
and (\ref{vilasi_spara_vino}) takes the well known form (see
\cite{Steph-et-al})%

\begin{equation}
g=\dfrac{1}{\sqrt{\left\vert \mu\right\vert }}\left(  d\mu^{2}+d\tilde{\mu
}^{2}\right)  +2\mu dp\left(  dq+Mdp\right)  \label{quella_buona}%
\end{equation}
with $M=h\left(  \mu,\tilde{\mu}\right)  /2-q$.

Hence, in order to find concrete Ricci flat metrics of the form
(\ref{quella_buona}) it is necessary to find exact solutions of
(\ref{pinkus_quasi_pallinus}) or, equivalently, equation (\ref{eqnostra})
which is obtained by (\ref{pinkus_quasi_pallinus}) through coordinate
transformation $\left\{  x=\mu+\tilde{\mu},\right.  \allowbreak\left.
y=\mu-\tilde{\mu}\right\}  $.

The most efficient way to do this consists in finding classical and higher
symmetries of such equations and using them to generate solutions (see next
section for a brief recall of the notion of symmetry of a system of PDE's; for
further details look at references therein).

Classical symmetries of (\ref{eqnostra}) have been already studied in our
previous paper \cite{CaMaPu05}. The aim of the present paper is to describe a
subalgebra of higher symmetries of (\ref{eqnostra}) by reducing it, via a
complex contact transformation $H$ (see section
\ref{sec.Symm.via.complex.trans}), to its hyperbolic analogue%
\begin{equation}
\mathcal{Y}_{ED}=\left\{  2\left(  \xi+\eta\right)  u_{\xi\eta}+u_{\xi
}+u_{\eta}=0\right\}  \text{,} \label{eqShem}%
\end{equation}
whose higher symmetries have been studied in detail in \cite{Shem93},
\cite{Shem94} and \cite{Shem95}. Of course, using a complex transformation
involves several problems of both geometrical and analytical kind. In
particular, one must complexify the original bundle $\pi:M\times
\mathbb{R}\rightarrow M$ and the corresponding jet bundles. Now, the
complexification of a real analytical manifold can be done in several ways,
all locally equivalent (\cite{Whitney-Bruhat}). However, we are interested in
preserving the jet bundle structure and, furthermore, in the possibility of
holomorphically extending real analytical functions to the \textit{whole}
complexified jet bundles. For these reasons, in the rest of the paper it will
be assumed that: 1) $M=\mathbb{R}^{2}$; 2) the only functions and differential
operators to be considered on the original real jet bundles are those which
rationally depend on base and jet variables. In fact, as we are mainly
interested in local coordinate expressions of infinitesimal symmetries of
(\ref{eqnostra}), the first assumption does not represent too severe a
restriction. As to the rationality assumption, it has been made to ensure the
holomorphic \textit{global} extendability. In particular, we construct a
vector space isomorphism between rational higher symmetries of equations
(\ref{eqShem}) and (\ref{eqnostra}).

\smallskip The paper is structured as follows.

In section \ref{sec.Preliminary} we give basic notions of the theory of
symmetries of differential equations, which we interpret as submanifolds of
jet bundles. Namely, we regard equation (\ref{eqnostra}) as a submanifold of
the second jet bundle $J^{2}\left(  \pi\right)  $. In this geometrical
setting, higher symmetries are vector fields on the infinite prolongation of
$\mathcal{E}_{ED}$ preserving its contact structure; in fact, such vector
fields can be identified with the corresponding generating functions, so that
we use the latter in concrete computations.

In section \ref{sec.Symm.via.complex.trans} we analyze the action
of complex transformation $H$ on a generic linear partial
differential equation $\mathcal{E}$ in two independent variables,
with rational coefficients. After having explicitly determined
prolongation $H$ and its most remarkable properties, we construct,
starting from it, a vector space isomorphism $\Theta$ between the
algebrae of rational higher symmetries of equation $\mathcal{E}$
and of $\ $transformed equation $\mathcal{Y}$ ; roughly speaking,
$\Theta$ maps any rational symmetry $\varphi$ of $\mathcal{Y}$
into the sum of the real and imaginary part of the pullback of
$\varphi$ along $H$ (theorem \ref{th.iniettiva}). By a completely
analogous reasoning one obtains a corresponding isomorphism
between the rational recursion operators of the two equations
(theorem \ref{isoRecur}).

In section \ref{sec.structure} we give concrete examples of computation,
taking into consideration the results of \cite{Shem93}, \cite{Shem94} and
\cite{Shem95}. We explicitly find symmetries of $\mathcal{E}_{ED}$ depending
on second derivatives, obtaining as a by-product contact symmetries already
found in \cite{CaMaPu05}. Finally, up to solutions of $\mathcal{E}_{ED}$, the
Lie algebra structure of rational higher symmetries of $\mathcal{E}_{ED}$ is
completely determined.

\section{Basic notions on symmetries}

\label{sec.Preliminary}

In this section we recall the basics about jet bundles and symmetries of PDE
(for further details see \cite{Many99},\cite{KrVe98},\cite{Olver01}%
,\cite{Sau89},\cite{Vin81},\cite{Vin88}).

Let $M$ be an $n$-dimensional smooth manifold and $\pi\colon E\rightarrow M$
be a vector bundle, $\dim E=n+m$. Let $\mathcal{U}\subset M$ be a neighborhood
of $M$ such that $\pi^{-1}(\mathcal{U})\simeq\mathcal{U}\times\mathbb{R}^{m}$
and let $(x_{\lambda},u^{i})$, $\lambda=1\dots n,\,i=1\dots m$, with
$(x_{\lambda})$ coordinates on $\mathcal{U}$, be the corresponding
trivialization. Then a local section of $\pi$ is locally given by $u^{i}%
=f^{i}(x_{1},x_{2},\ldots,x_{n})$. We shall denote by $\Gamma(\pi)$ the
$C^{\infty}\left(  M\right)  $-module of local sections of $\pi$.

\smallskip

Two local sections $s$ and $\tilde{s}$ of $\pi$ are said to be \emph{$r$%
-contact equivalent} at the point $x\in M$ if their Taylor expansions at this
point coincide up to order $r$. This is an equivalence relation, and we shall
denote by $[s]_{x}^{r}$ an equivalence class. The set $J^{r}(\pi)$ of all the
equivalence classes $[s]_{x}^{r}$ is called the \emph{jet bundle} of order $r$
and it has a natural vector bundle structure. A chart $(x_{\lambda
},u_{\mathbf{\sigma}}^{i})$ on $J^{r}(\pi)$ is defined by $u_{\sigma}%
^{i}(\left[  s]_{x}^{r}\right)  =\frac{\partial^{\lvert\sigma\rvert}f^{i}%
}{\partial x_{\sigma}}(x)$, where $\sigma=\left(  \sigma_{1},\sigma_{2}%
,\dots,\sigma_{n}\right)  $ with $|\sigma|\overset{\text{def}}{=}\sum
\sigma_{i}\leq r$ and $0\leq\sigma_{i}\leq n$, is a multi-index and
$\frac{\partial^{\lvert\sigma\rvert}}{\partial x_{\sigma}}$ stands for
$\frac{\partial^{\lvert\sigma\rvert}}{\partial x_{1}^{\sigma_{1}}%
\cdots\partial x_{n}^{\sigma_{n}}}$.

\bigskip

We have the following natural maps:

\begin{enumerate}
\item  the embedding $j_{r}s\colon M\rightarrow J^{r}(\pi),\,\,x\mapsto\lbrack
s]_{x}^{r}$\thinspace,

\item  the projections $\pi_{k,h}\colon J^{k}(\pi)\rightarrow J^{h}%
(\pi),\,[s]_{x}^{k}\mapsto\lbrack s]_{x}^{h}\,\quad k\geq h$,

\item  The base projections $\pi_{r}:J^{r}(\pi)\rightarrow M$, \ $[s]_{x}^{r}%
$\thinspace$\mapsto x$.
\end{enumerate}

Note that there is a natural bijection between smooth functions on
$J^{r}\left(  \pi\right)  $ and scalar $r$-th order differential operators on
$\Gamma\left(  \pi\right)  $. Namely, with each $F\in C^{\infty}\left(
J^{r}(\pi)\right)  $ one can associate the following operator
\[
\Delta_{F}:\Gamma\left(  \pi\right)  \rightarrow C^{\infty}\left(  M\right)
,\,\,\,s\mapsto F\circ j_{r}s\text{, \ \ \ }s\in\Gamma\left(  \pi\right)  .
\]

The \emph{contact plane} $\mathcal{C}_{\theta}^{r}$ at the point $\theta\in
J^{r}(\pi)$ is the span of the planes $T_{\theta}\left(  j_{r}s(M)\right)  $,
with $s\in\Gamma\left(  \pi\right)  $ varying among sections whose $r$-jet at
$\pi_{r}\left(  \theta\right)  $ coincides with $\theta$. We have the
\emph{contact distribution} $\theta\mapsto\mathcal{C}_{\theta}^{r}$ on
$J^{r}(\pi)$. A diffeomorphism of $J^{r}(\pi)$ is called a \emph{contact
transformation} if it is a symmetry of the contact distribution (i.e. if it
preserves contact planes). A vector field on $J^{r}(\pi)$ whose flow consists
of contact transformations is called a \emph{contact field}. We note that a
point $\theta=[s]_{x}^{r+1}$ of $J^{r+1}(\pi)$ is completely characterized by
$T_{\pi_{r+1,r}(\theta)}\left(  j_{r}s(M)\right)  $. Then we can lift a
contact transformation $G$ of $J^{r}(\pi)$ to a contact transformation
$G^{(1)}$ of $J^{r+1}(\pi)$ by considering $G_{\ast}(T_{\pi_{r+1,r}(\theta
)}(j_{r}s(M)))$. Of course we can lift contact fields by lifting their local
flows. According to a classical result by Lie and Baecklund, any contact
transformation is the lifting: 1) of a first order contact transformation if
$m=\operatorname*{rank}\pi=1$; 2) of a diffeomorphism of $J^{0}\left(
\pi\right)  =E$ if $m>1$. An analogous result holds for contact fields.

\smallskip A \emph{differential equation} $\mathcal{E}$ of order $r$ is a
submanifold of $J^{r}(\pi)$. A \emph{linear} equation is a linear subbundle of
$J^{r}(\pi)\rightarrow M$. A (local) \emph{solution} of $\mathcal{E}$ is a
section $s$ of $\pi$ such that $j_{r}s(M)\subset\mathcal{E}$. The
\emph{1-prolongation} $\mathcal{E}^{1}$ of the equation $\mathcal{E}$ is the
set of first order ``differential consequences''\ of $\mathcal{E}$.
Geometrically:
\[
\mathcal{E}^{1}=\{[s]_{x}^{r+1}\in J^{r+1}(\pi)\,|\,s\in\Gamma(\pi
),~[s]_{x}^{r}\in\mathcal{E},\,\,T_{[s]_{x}^{r}}(j_{r}s(M))\subset
T_{[s]_{x}^{r}}\mathcal{E}\}.
\]
By iteration we can define the $l$-prolongation $\mathcal{E}^{l}$. Locally, if
the equation $\mathcal{E}$ is described by $\left\{  F^{i}=0\right\}  $, with
$F^{i}\in C^{\infty}\left(  J^{r}\left(  \pi\right)  \right)  $, then
$\mathcal{E}^{l}$ is described by $\left\{  D_{\sigma}\left(  F^{i}\right)
=0\right\}  $ with $0\leq|{\sigma}|\leq l$, where $D_{\sigma}=D_{\sigma_{1}%
}\circ D_{\sigma_{2}}\circ\dots\circ D_{\sigma_{n}}$ and $D_{\lambda}$ are
\emph{total derivatives}:%
\[
D_{\lambda}=\frac{\partial}{\partial x_{\lambda}}+\sum_{j,\sigma}%
u_{\sigma,\lambda}^{j}\frac{\partial}{\partial u_{\sigma}^{j}}.
\]

\smallskip Note that the above definitions of $r$-contact equivalence and
$r$-th order jet space make sense even in the case $r=\infty$. Obviously,
$J^{\infty}\left(  \pi\right)  $ is \textit{not} a finite dimensional smooth
manifold (its points are sequences of the form $\left\{  \theta_{r}\right\}
,\, r\in\mathbb{N}_{0}$, with $\theta_{r}\in J^{r}\left(  \pi\right)  $ and
$\pi_{r,r-1}\left(  \theta_{r}\right)  =\theta_{r-1}$ ). However, a very rich
differential calculus can be developed on it, making it an extremely useful
tool in symmetry analysis of PDE's as well as in many other fields. Here we
limit ourselves to recall just a few basic facts about the differential
structure on $J^{\infty}\left(  \pi\right)  $ (for further details see
\cite{Many99}):

\smallskip By definition, smooth functions on $J^{\infty}\left(  \pi\right)  $
are pullbacks of smooth functions on finite order jet spaces along projections
$\pi_{\infty,k}$. Thus, $C^{\infty}\left(  J^{\infty}\left(  \pi\right)
\right)  $ is a filtered algebra (the degree being the jet order of the
pullbacked function). Consequently, vector fields on $J^{\infty}\left(
\pi\right)  $ are defined as derivations $X:C^{\infty}\left(  J^{\infty
}\left(  \pi\right)  \right)  \rightarrow$ $C^{\infty}\left(  J^{\infty
}\left(  \pi\right)  \right)  $ such that $\deg X\left(  f\right)  -\deg f$ is
a constant integer depending only on $X$. Vector fields on $J^{\infty}(\pi)$
do not admit, generally, a flow, even locally. For instance, $D_{\lambda}$ is
a vector field on $J^{\infty}(\pi)$ with degree $1$. A tangent vector at a
point $\theta=\left\{  \theta_{r}\right\}  \in J^{\infty}\left(  \pi\right)  $
is a sequence $\xi=\left\{  \xi_{r}\right\}  $ such that $\xi_{r}\in
T_{\theta_{r}}J^{r}\left(  \pi\right)  $ and $d_{\theta_{r}}\pi_{r,r-1}\left(
\xi_{r}\right)  =\xi_{r-1}$. The contact plane $\mathcal{C}_{\theta}$ is the
sequence $\{\mathcal{C}_{\theta_{r}}^{r}\}$. Contact distribution
$\theta\longmapsto\mathcal{C}_{\theta}$ on $J^{\infty}(\pi)$ is $n$%
-dimensional (it is spanned by total derivatives $\{D_{\lambda}\}_{1\leq
\lambda\leq n}$) and formally integrable, in the sense that its generators
satisfy Frobenius conditions. A vector field on $J^{\infty}(\pi)$ lying in the
contact distribution $\mathcal{C}$ is called \emph{trivial} as it is tangent
to all integral manifolds of $\mathcal{C}$. Any contact field $X$ on
$J^{\infty}(\pi)$ can be splitted in a vertical and a trivial part. More
precisely we have that:%
\[
X=X_{\varphi}+T
\]
where%
\begin{equation}
X_{\varphi}=\sum_{j,\sigma}D_{\sigma}(\varphi^{j})\frac{\partial}{\partial
u_{\sigma}^{j}}\text{,} \label{evolut_field}%
\end{equation}
with $\varphi=\left(  \varphi_{1},\ldots,\varphi_{m}\right)  $, $\varphi
_{j}\in C^{\infty}\left(  J^{\infty}\left(  \pi\right)  \right)  $ and $T$ is
a trivial vector field. It can be proved that fields of the form
(\ref{evolut_field}), called \emph{evolutionary vector fields}, are the only
vertical contact fields on $J^{\infty}\left(  \pi\right)  $; vector function
$\varphi$ is called the \emph{generating section} of $X_{\varphi}$ (also know
as characteristic of a contact field, \cite{Olver01}). Evolutionary vector
fields form a Lie subalgebra, isomorphic to the algebra of generating sections
with respect to \emph{Jacobi bracket}%
\begin{equation}
\left\{  \varphi,\psi\right\}  \overset{\text{def}}{=}X_{\varphi}\left(
\psi\right)  -X_{\psi}\left(  \varphi\right)  \label{jacobrack}%
\end{equation}

\smallskip A \emph{classical symmetry} of $\mathcal{E}$ is a contact field on
$J^{r}(\pi)$ tangent to $\mathcal{E}$. If, in particular, it is a lift of a
vector field on $E$, then it is called a \emph{point symmetry}. A contact
field on $J^{\infty}(\pi)$ tangent to $\mathcal{E}^{\infty}$ is called an
\emph{external higher symmetry}. A vector field on $\mathcal{E}^{\infty}$
which preserves the contact distribution induced on $\mathcal{E}^{\infty}$ is
called an \emph{internal higher symmetry}.

\smallskip Now we are interested in non-trivial symmetries, that is symmetries
of the form $X_{\varphi}$. Locally, if the equation $\mathcal{E}$ is described
by $\left\{  F^{i}=0\right\}  $, with $F^{i}\in C^{\infty}\left(  J^{r}\left(
\pi\right)  \right)  $, then the vector field $X_{\varphi}$ is a symmetry of
$\mathcal{E}$ iff $X_{\varphi}(F^{i})|_{\mathcal{E}^{\infty}}=0.$ If we define
matrix operator $\ell_{F}$ by
\[
\ell_{F}\left(  \varphi\right)  \overset{\text{def}}{=}\left\|  X_{\varphi
^{j}}\left(  F^{i}\right)  \right\|
\]
we have that $\varphi$ is an external higher symmetry of $\mathcal{E}$ if and
only if
\begin{equation}
\left(  \ell_{F}(\varphi)\right)  |_{\mathcal{E}^{\infty}}=0\text{.}
\label{eq.of.symmetries}%
\end{equation}

The operator $\ell_{\mathcal{E}}=\ell_{F}|_{\mathcal{E}^{\infty}}$ is called
\emph{universal linearization} of $\mathcal{E}$. Locally we have that
$\varphi$ is an external higher symmetry if
\begin{equation}
\ell_{\mathcal{E}}(\bar{\varphi})=\sum_{j,\sigma}\frac{\partial F^{i}%
}{\partial u_{\sigma}^{j}}\bar{D}_{\sigma}(\bar{\varphi}^{j})=0
\label{cazzacchione}%
\end{equation}
where the bar denotes the restriction to $\mathcal{E}^{\infty}$.

It is easy to realize that any external higher symmetry restricts to an
internal higher symmetry. The converse is also true: each internal higher
symmetry can be obtained by restricting on $\mathcal{E}^{\infty}$ some
external one. For this reason we shall not distinguish them, and we shall call
them simply higher symmetries. Then we shall denote by $\operatorname*{Sym}%
(\mathcal{E})$ the algebra of (non-trivial) higher symmetries of $\mathcal{E}$.

A vector valued operator $\Delta$ acting on vector functions on $J^{\infty
}(\pi)$ is called \emph{$\mathcal{C}$-differential} if its restriction to
$\mathcal{E}^{\infty}$ is well defined for any differential equation
$\mathcal{E}$. In local coordinates, $\Delta$ reads
\[
\Delta=\left\|  \sum_{\sigma}a_{ij}^{\sigma}D_{\sigma}\right\|  \,,\quad
\text{where}\quad a_{ij}^{\sigma}\in C^{\infty}(J^{\infty}(\pi)).
\]
An example of $\mathcal{C}$-differential operator is given by $\ell_{F}$.

Finally, a \emph{recursion operator} $\Re\in\operatorname*{Rec}(\mathcal{E})$
for a differential equation $\mathcal{E}$ is a linear $\mathcal{C}%
$-differential operator which maps $\operatorname*{Sym}(\mathcal{E})$ into itself.

\section{Symmetries and recursion operators of linear equations by complex transformations}

\label{sec.Symm.via.complex.trans}

In this section it will be shown how do symmetries of a scalar linear PDE in
two independent variables transform under a complex point transformation. As
we are mainly interested in local coordinate expressions of infinitesimal
symmetries, we can safely assume that the bundle of independent and dependent
variables is the trivial bundle $\pi:\mathbb{R}^{2}\times\mathbb{R}%
\rightarrow\mathbb{R}^{2}$.

Consider the $r$-th order scalar PDE
\begin{equation}
\mathcal{E}=\left\{  F\left(  x,y,u,u_{x},u_{y},...,u_{hx,ky},...,u_{ry}%
\right)  =0\right\}  \label{FirstEq}%
\end{equation}
where $\left(  x,y,u\right)  $ are standard coordinates on $\mathbb{R}%
^{2}\times\mathbb{R}$, $u_{hx,ky}\overset{\text{def}}{=}u_{\underset
{\text{h-times}}{\underbrace{xx...x}}\underset{\text{k-times}}{\underbrace
{yy...y}}}$, $h+k\leq r$, and $F$ is assumed to be linear in $u,u_{x}%
,u_{y},...,u_{hx,ky},...,u_{ry}$ with coefficients which are rational
functions in $x,y$ (the reason of this assumption has been given in the
introduction). We want to express $\mathcal{E\ }$in terms of complex conjugate
variables $z,\overline{z}$, with $z=x+iy$. The complex analogues of jet
variables are defined as follows. Recall that to $z,\overline{z}$ one can
associate formal partial derivatives:
\begin{equation}
\frac{\partial}{\partial z}\overset{\text{def}}{=}\frac{1}{2}\left(
\frac{\partial}{\partial x}-i\frac{\partial}{\partial y}\right)  \text{,
\ \ \ \ \ \ \ \ \ \ \ \ \ \ }\frac{\partial}{\partial\overline{z}}%
\overset{\text{def}}{=}\frac{1}{2}\left(  \frac{\partial}{\partial x}%
+i\frac{\partial}{\partial y}\right)  \label{Wirty}%
\end{equation}
These are vector fields on the complexified tangent bundle $T_{\mathbb{C}%
}\left(  \mathbb{R}^{2}\right)  \simeq\mathbb{R}^{2}\times\mathbb{C}^{2}$. Let
also $\pi_{\mathbb{C}}:\mathbb{R}^{2}\times\mathbb{C}\rightarrow\mathbb{R}%
^{2}$ be the complexified bundle of $\pi$. One can immediately extend to it
the jet bundle construction of section \ref{sec.Preliminary}. Let
$s=s_{1}+is_{2}$ be a section of $\pi_{\mathbb{C}}$ and $\theta=\left[
s\right]  _{a}^{k}$ be its $k$-jet at point $a\in\mathbb{R}^{2}$. Then its
complex conjugate is, by definition, $\overline{\theta}=\left[  \overline
{s}\right]  _{a}^{k}$ and its jet coordinates $\overline{u}_{rx,py}$ are
complex conjugate to those $u_{rx,py}$ of $\theta$.

Starting from (\ref{Wirty}) one defines by iterated composition higher order
partial derivatives of sections of \ $\pi_{\mathbb{C}}$ with respect to base
``coordinates'' $z$, $\overline{z}$. This allows to define complex jet
coordinates of $\theta$ as%
\begin{equation}
u_{hz,\,l\overline{z}}\overset{\text{def}}{=}\frac{\partial^{h+l}s}{\partial
z^{h}\partial\overline{z}^{l}}\left(  a\right)  \text{,}
\label{compl_jet_definition}%
\end{equation}
for $0\leq h+l\leq k$.

\begin{proposition}
Let $s\in\Gamma\left(  \pi_{\mathbb{C}}\right)  $. Then%
\begin{equation}
\overline{\left(  \frac{\partial^{h+l}s}{\partial z^{h}\partial\overline
{z}^{l}}\right)  }=\frac{\partial^{h+l}\overline{s}}{\partial z^{k}%
\partial\overline{z}^{l}}\text{,} \label{conjug_derive}%
\end{equation}
for any $h,l\in\mathbb{N}$.
\end{proposition}

\begin{proof}
It easily follows from (\ref{Wirty}) by induction .
\end{proof}

An immediate consequence of the previous proposition is the following relation
between $k$-th order complex jet variables of \textit{real} jets
($\theta=\overline{\theta}$):%
\begin{equation}
\overline{u_{\left(  k-r\right)  z,r\overline{z}}}=u_{rz,\left(  k-r\right)
\overline{z}}\text{,} \label{conjug_u}%
\end{equation}

Analogously, one can define also total derivatives with respect to $z$,
$\overline{z}$. First order total derivatives are, by definition
\begin{equation}
D_{z}=\frac{1}{2}\left(  D_{x}-iD_{y}\right)  \text{, \ \ \ \ \ \ \ \ \ }%
D_{\overline{z}}=\frac{1}{2}\left(  D_{x}+iD_{y}\right)  \label{WirtyTotal}%
\end{equation}
and higher order total derivatives are obtained by (\ref{WirtyTotal}) via
iterated composition.

\bigskip

Complex jet variables are related to real ones by linear relation:
\begin{equation}
\left(  z,\overline{z},u,u_{z},u_{\overline{z}},\ldots\right)  ^{T}%
=H\cdot\left(  x,y,u,u_{x},u_{y},\ldots\right)  ^{T}\text{,}
\label{change_coordinates}%
\end{equation}
where $H$ is the (infinite) block matrix%
\[
H=\left(
\begin{array}
[c]{lllllll}%
P &  &  &  &  &  & \\
& 1 &  &  &  &  & \\
&  & P^{(1)} &  &  &  & \\
&  &  & P^{(2)} &  &  & \\
&  &  &  & \ddots &  & \\
&  &  &  &  &  P^{(k)} & \\
&  &  &  &  &  & \ddots
\end{array}
\right)  \text{,}%
\]
with%
\begin{equation}
P=\left(
\begin{array}
[c]{rr}%
1 & i\\
1 & -i
\end{array}
\right)  \label{eq.P}%
\end{equation}
and $P^{(k)}$ defined by%
\begin{equation}
V^{\left(  k\right)  }=P^{\left(  k\right)  }U^{\left(  k\right)  }\text{,}
\label{VPU}%
\end{equation}
with $V^{\left(  k\right)  }=\left(  u_{k{z}},u_{(k-1)z,\overline{z}%
},...,u_{k\overline{z}}\right)  ^{T}$, $U^{\left(  k\right)  }=\left(
u_{kx},u_{\left(  k-1\right)  x,y},...,u_{ky}\right)  ^{T}$, $P^{\left(
k\right)  }=\left\Vert p_{rs}^{k}\right\Vert _{r,s=0,...,k}$. Any block is
obtained by the previous one via recursive formulae (\ref{recurr}) below.
Therefore, the whole matrix $H$ is determined by the first block $P$: in fact,
$H$ is just the contact prolongation of coordinate transformation $H_{0}$ on
$J^{0}\left(  \pi_{\mathbb{C}}\right)  $, given by:
\begin{equation}
\left(  z,\overline{z}\right)  ^{T}=P\cdot\left(  x,y\right)  ^{T},u=u.
\label{eq.H0}%
\end{equation}

In next section we will explicitly determine coefficients of
matrix $H$ and show some remarkable properties of it.

\subsection{Explicit formulae for matrix $H$}

It is clear that, for any $k\in\mathbb{N}$ and $r=0,1,\cdots,k$, $D_{\left(
k-r\right)  z,r\overline{z}}$ is a linear combination of standard total
derivatives of order $k$:%
\begin{equation}
D_{\left(  k-r\right)  z,r\overline{z}}=\sum_{q=0}^{r}p_{rq}^{k}D_{\left(
k-q\right)  x,qy}. \label{matrix1}%
\end{equation}
Consequently, keeping in mind that, as in the real case, complex jet variables
can be expressed by formula
\begin{equation}
u_{hz,\,k\overline{z}}=D_{hz,\,k\overline{z}}\left(  u\right)  \text{,}
\label{jet_comples_D}%
\end{equation}
with $D_{hz,k\overline{z}}=D_{z}^{h}\circ D_{\overline{z}}^{k}$, we have that
\begin{equation}
u_{\left(  k-r\right)  z,r\overline{z}}=\sum_{q=0}^{k}p_{rq}^{k}\,u_{\left(
k-q\right)  x,qy} \label{def_pkrs}%
\end{equation}
In order to compute coefficient $p_{rs}^{k}$, let us apply both hands of
(\ref{matrix1}) to the function $x^{k-s}y^{s}$. Such a function depends only
on base variables, so its total derivatives coincide with the corresponding
partial ones. Hence, the only non zero term in the right hand side of
(\ref{matrix1}) applied to $x^{k-s}y^{s}$ is that with $q=s$, i.e.%
\[
D_{\left(  k-r\right)  z,r\overline{z}}\left(  x^{k-s}y^{s}\right)  =s!\left(
k-s\right)  !p_{rs}^{k}%
\]
Hence%
\begin{equation}
p_{rs}^{k}=\frac{1}{s!\left(  k-s\right)  !}D_{\left(  k-r\right)
z,r\overline{z}}\left(  x^{k-s}y^{s}\right)  \label{matrix2}%
\end{equation}
To compute the total derivative in (\ref{matrix2}) it is convenient to express
$x^{k-s}y^{s}$ in terms of $z$ and $\overline{z}$:%
\begin{align*}
x^{k-s}y^{s}  &  =\frac{i^{s}}{2^{k}}\left(  z+\overline{z}\right)
^{k-s}\left(  \overline{z}-z\right)  ^{s}\\
&  =\frac{i^{s}}{2^{k}}\sum_{\alpha=0}^{k-s}\sum_{\beta=0}^{s}\left(
-1\right)  ^{\beta}\binom{k-s}{\alpha}\binom{s}{\beta}z^{\alpha+\beta
}\overline{z}^{k-\alpha-\beta}%
\end{align*}
Now, following the same argument as above, the only terms in this double sum
on which $D_{\left(  k-r\right)  z,r\overline{z}}$ does not vanish are those
corresponding to the monomial $z^{k-r}\overline{z}^{r}$, i.e.%
\begin{align*}
&  \frac{i^{s}}{2^{k}}\sum_{\alpha+\beta=k-r}\left(  -1\right)  ^{\beta}%
\binom{k-s}{\alpha}\binom{s}{\beta}z^{\alpha+\beta}\overline{z}^{k-\alpha
-\beta}\\
&  =\left(  -1\right)  ^{k-r}\frac{i^{s}}{2^{k}}\sum_{\alpha=M\left(
k,r,s\right)  }^{m\left(  k,r,s\right)  }\left(  -1\right)  ^{\alpha}%
\binom{k-s}{\alpha}\binom{s}{k-r-\alpha}z^{k-r}\overline{z}^{r}\text{,}%
\end{align*}
with $M\left(  k,r,s\right)  =\max\left(  0,k-r-s\right)  $, \ $m\left(
k,r,s\right)  =\min\left(  k-r,k-s\right)  $. Hence, (\ref{matrix2}) can be
rewritten as follows:%
\begin{align}
p_{rs}^{k}  &  =\left(  -1\right)  ^{k-r}\frac{r!\left(  k-r\right)  !}%
{2^{k}s!\left(  k-s\right)  !}\,i^{s}\sum_{\alpha=M\left(  k,r,s\right)
}^{m\left(  k,r,s\right)  }\left(  -1\right)  ^{\alpha}\binom{k-s}{\alpha
}\binom{s}{k-r-\alpha}\label{pkrs}\\
&  =\left(  -1\right)  ^{k-r}\frac{r!\left(  k-r\right)  !}{2^{k}}\,i^{s}%
\sum_{\alpha=M\left(  k,r,s\right)  }^{m\left(  k,r,s\right)  }\frac{\left(
-1\right)  ^{\alpha}}{\alpha!\left(  k-s-\alpha\right)  !\left(
k-r-\alpha\right)  !\left(  r+s+\alpha-k\right)  !}\nonumber
\end{align}

\subsubsection{Some remarkable properties of matrix
$H\label{sec_remarkable_properties}$}

\begin{itemize}
\item  It is obvious that, for any $k\in\mathbf{N}$, the block $P^{\left(
k\right)  }$ is invertible (one must be able to express $k$-th order standard
jet variables in term of complex ones). It can be proved that the inverse is
$P^{\left(  k\right)  ^{-1}}=\left\|  q_{rs}^{k}\right\|  $, with%
\[
q_{rs}^{k}=2^{k}i^{r+s}p_{rs}^{k}%
\]

\item  For each block $P^{\left(  k\right)  }$ the following symmetry property
holds%
\begin{equation}
\overline{p_{rq}^{k}}=p_{k-r,q}^{k}\text{,} \label{symmetry_P_k}%
\end{equation}
for $r,q=0,\ldots,k$. This is an immediate consequence of (\ref{def_pkrs}) and
(\ref{conjug_u}).

\item  For any $k$, $\overline{P^{\left(  k\right)  }}\cdot P^{\left(
k\right)  ^{-1}}$ is real. In fact, considering, as above, only real sections
of $\pi_{\mathbb{C}}$ and taking the complex conjugate of both hands of
(\ref{VPU}), one gets%
\begin{equation}
\overline{V^{\left(  k\right)  }}=\overline{P^{\left(  k\right)  }}U^{\left(
k\right)  }=\overline{P^{\left(  k\right)  }}P^{\left(  k\right)  ^{-1}%
}V^{\left(  k\right)  } \label{real1}%
\end{equation}
But, keeping in mind (\ref{conjug_u}), one also gets%
\begin{equation}
\overline{V^{\left(  k\right)  }}=AV^{\left(  k\right)  }\text{,}
\label{real2}%
\end{equation}
where $A=\left\|  \delta_{i,k-j}\right\|  _{i,j=0,\ldots,k}$. This, together
with (\ref{real1}), implies that $\overline{P^{\left(  k\right)  }}P^{\left(
k\right)  ^{-1}}=A$

\item  The following recurrence formulae hold:%
\begin{align}
p_{r,s}^{k+1}  &  =\frac{1}{2}\left(  p_{r,s}^{k}-ip_{r,s-1}^{k}\right)
\text{ \ \ \ \ \ \ \ for }r\leq k\label{recurr}\\
p_{k+1,s}^{k+1}  &  =\frac{1}{2}\left(  p_{k,s}^{k}+ip_{k,s-1}^{k}\right)
\text{,}\nonumber
\end{align}
where we pose $p_{r,-1}^{k}=p_{r,k+1}^{k}=0$ \ for $r=0,1,\cdots k$. Such
formulae are obtained from relations $u_{\left(  k+1-r\right)  z,r\overline
{z}}=D_{z}\left(  u_{\left(  k-r\right)  z,r\overline{z}}\right)  $,
$u_{\left(  k+1\right)  \overline{z}}=D_{\overline{z}}\left(  u_{k\overline
{z}}\right)  $ together with ($\ref{WirtyTotal}$) e (\ref{def_pkrs}).
\end{itemize}

\subsection{The transformed equation and its symmetries}

\label{sec.transf.equ.and.its.symm}

Before proving the main result of this section, i.e. theorem \ref{zaza6}, some
remarks about linear transformation (\ref{change_coordinates}) are necessary.

First of all, (\ref{change_coordinates}) can be extended to complex values of
variables $\left(  x,y,u,u_{x},u_{y},\ldots\right)  $; in other words, $H$ can
be viewed as a contact transformation
\begin{equation}
\left(  \xi,\eta,u,u_{\xi},u_{\eta},\ldots\right)  ^{T}=H\cdot\left(
x,y,u,u_{x},u_{y},\ldots\right)  ^{T} \label{trasformaledetta}%
\end{equation}
of a complex jet space defined as follows.

Recall that a complexification of a real analytic manifold $N$ is a complex
manifold $\widehat{N}$ together with a real analytical embedding
$\alpha:N\rightarrow\widehat{N}$ and an involutive antiholomorphism
$\chi:\widehat{N}\rightarrow\widehat{N}$ such that $\chi\circ\alpha=\alpha$.
As in the previous section, let $\pi$ be the trivial bundle $\mathbb{R}%
^{2}\times\mathbb{R\rightarrow R}^{2}$. Below we construct a complexification
of $J^{\infty}\left(  \pi\right)  $ which is the right domain for
transformation (\ref{trasformaledetta}).

Let $\widehat{\pi}$ be the trivial holomorphic bundle $\mathbb{C}^{2}%
\times\mathbb{C\rightarrow C}^{2}$. $\widehat{\pi}$ is a natural
complexification of $\pi$ and transformation (\ref{eq.H0}) can be extended to
the following automorphism of $\widehat{\pi}$:%
\[
\left(  \xi,\eta\right)  ^{T}=P\cdot\left(  x,y\right)  ^{T}\text{,
\ \ }u=u\text{,}%
\]
with $\left(  x,y,u\right)  \in\mathbb{C}^{2}\times\mathbb{C}$. It is clear
that the jet bundle construction described in section \ref{sec.Preliminary}
can be repeated word by word for holomorphic jet bundles. Thus, starting from
$\widehat{\pi}$ one gets a sequence of jet bundles $J^{k}\left(  \widehat{\pi
}\right)  $ whose inverse limit we denote by $J^{\infty}\left(  \widehat{\pi
}\right)  $. It is easily seen that, for any $k\in\mathbb{N}$, $J^{k}\left(
\widehat{\pi}\right)  $ is a complexification of $J^{k}\left(  \pi\right)  $.
In fact, let $\iota_{k}:J^{k}\left(  \pi\right)  \rightarrow J^{k}\left(
\widehat{\pi}\right)  $ be defined in the following way. For any $s\in
\Gamma\left(  \pi\right)  $, $a\in\mathbb{R}^{2}$, denote by $P_{s,a}%
^{k}:\mathbb{R}^{2}\rightarrow\mathbb{R}$ the $k$-th order Taylor polynomial
of $s$ at $a$. Then%
\[
\iota_{k}\left(  \left[  s\right]  _{a}^{k}\right)  \overset{\text{def}}%
{=}\left[  \widehat{P}_{s,a}^{k}\right]  _{a}^{k}\text{,}%
\]
where $\widehat{P}_{s,a}^{k}:\mathbb{C}^{2}\rightarrow\mathbb{C}$ is the
complex extension of $P_{s,a}^{k}$. Furthermore, let $\sigma_{k}:J^{k}\left(
\widehat{\pi}\right)  \rightarrow J^{k}\left(  \widehat{\pi}\right)  $ be the
map $\left[  s\right]  _{a}^{k}\mapsto\left[  \overline{s}\right]  _{a}^{k}$.
The triple $\left(  J^{k}\left(  \widehat{\pi}\right)  ,\iota_{k},\sigma
_{k}\right)  $ is the required complexification of $J^{k}\left(  \pi\right)
$. The inverse limit $J^{\infty}\left(  \widehat{\pi}\right)  $ is a
complexification of $J^{\infty}\left(  \pi\right)  $ in the following sense.
Let $\iota:J^{\infty}\left(  \pi\right)  \rightarrow J^{\infty}\left(
\widehat{\pi}\right)  $ be the map defined by%
\[
\iota\left(  \theta\right)  =\left\{  \iota_{k}\left(  \theta_{k}\right)
\right\}  _{k\in\mathbb{N}_{0}}\text{,}%
\]
with $\theta=\left\{  \theta_{k}\right\}  _{k\in\mathbb{N}_{0}}$. Analogously,
the conjugate map $\sigma:J^{\infty}\left(  \widehat{\pi}\right)  \rightarrow
J^{\infty}\left(  \widehat{\pi}\right)  $ is defined as the inverse limit of
$\left\{  \sigma_{k}\right\}  _{k\in\mathbb{N}_{0}}$. So, by the phrase
``$J^{\infty}\left(  \widehat{\pi}\right)  $ is a complexification of
$J^{\infty}\left(  \pi\right)  $''\ we simply mean that we consider
$J^{\infty}\left(  \widehat{\pi}\right)  $ together with maps $\iota$ and
$\sigma$. We stress that, in accordance with the real case (see section
\ref{sec.Preliminary}), one must consider as holomorphic functions on
$J^{\infty}\left(  \widehat{\pi}\right)  $ only the pullbacks of holomorphic
functions on finite order jet bundles $J^{k}\left(  \widehat{\pi}\right)  $.

\smallskip We introduce the following definitions and notations. Denote by:

\begin{itemize}
\item $\Omega\left(  J^{\infty}\left(  \pi\right)  \right)  $ the algebra of
real analytic functions on $J^{\infty}\left(  \pi\right)  $, i.e. (see section
\ref{sec.Preliminary}) pullbacks of analytic functions on finite order jet
bundles along projections $\pi_{\infty,k}$

\item $\Omega\left(  J^{\infty}\left(  \widehat{\pi}\right)  \right)  $ the
algebra of complex analytic functions on $J^{\infty}\left(  \widehat{\pi
}\right)  $, in the same sense as the point above.

\item $\Omega_{\mathbb{C}}\left(  J^{\infty}\left(  \pi\right)  \right)
\overset{\text{def}}{=}\left\{  f_{1}+if_{2}\,|\,f_{1},f_{2}\in\Omega\left(
J^{\infty}\left(  \pi\right)  \right)  \right\}  $.
\end{itemize}

\smallskip Under the action of $H$ equation (\ref{FirstEq}) is transformed
into equation
\begin{equation}
\mathcal{Y}=\left\{  \widetilde{F}\left(  \xi,\eta,u,\cdots,u_{r\eta}\right)
=0\right\}  \text{,} \label{SecondEq}%
\end{equation}
with $\widetilde{F}=\widehat{F}\circ H^{-1}\circ\iota$. Here by $\widehat{F}$
we mean the holomorphic extension of $F$ to $J^{\infty}\left(  \widehat{\pi
}\right)  $. Note that such an extension is well defined due to the particular
form of $F$.

As we noted in the introduction, analytic functions on $J^{\infty}(\pi)$ are
not generally holomorphically extendable on the complexification. Due to this
reason, below we shall restrict our attention to rational functions of
$x,y,u,u_{x},u_{y},...,u_{hx,ky},...$. In other words, we consider, instead of
$\Omega\left(  J^{\infty}\left(  \widehat{\pi}\right)  \right)  $, the
subalgebra $\mathcal{R}\left(  J^{\infty}\left(  \widehat{\pi}\right)
\right)  $ of complex rational functions of (a finite number of) jet
variables. Analogously, we denote by $\mathcal{R}\left(  J^{\infty}\left(
\pi\right)  \right)  \subset\Omega\left(  J^{\infty}\left(  \pi\right)
\right)  $ the subalgebra of real rational functions on $J^{\infty}\left(
\pi\right)  $ and by $\mathcal{R}_{\mathbb{C}}\left(  J^{\infty}\left(
\pi\right)  \right)  \subset\Omega_{\mathbb{C}}\left(  J^{\infty}\left(
\pi\right)  \right)  $ the algebra of functions of the form $f_{1}+if_{2}$,
with $f_{i}\in\mathcal{R}\left(  J^{\infty}\left(  \pi\right)  \right)  $.

With these restrictions, the holomorphic extension operator%
\[
\rho:\mathcal{R}_{\mathbb{C}}\left(  J^{\infty}\left(  \pi\right)  \right)
\rightarrow\mathcal{R}\left(  J^{\infty}\left(  \widehat{\pi}\right)
\right)
\]
is well defined (as above, we will occasionally denote $\rho\left(  f\right)
$ by $\widehat{f}$).

Below we will need to consider also holomorphic extension of differential operators.

Obviously, any linear differential operator on $\mathcal{R}\left(  J^{\infty
}\left(  \pi\right)  \right)  $ with rational coefficients can be extended by
$\mathbb{C}$-linearity to an operator on $\mathcal{R}_{\mathbb{C}}\left(
J^{\infty}\left(  \pi\right)  \right)  $.

With any differential operator $\Delta:\mathcal{R}_{\mathbb{C}}\left(
J^{\infty}\left(  \pi\right)  \right)  \rightarrow\mathcal{R}_{\mathbb{C}%
}\left(  J^{\infty}\left(  \pi\right)  \right)  $ one associates its image
$H\left(  \Delta\right)  :\mathcal{R}_{\mathbb{C}}\left(  J^{\infty}\left(
\pi\right)  \right)  \rightarrow\mathcal{R}_{\mathbb{C}}\left(  J^{\infty
}\left(  \pi\right)  \right)  $, defined as follows%
\begin{equation}
H\left(  \Delta\right)  \overset{\text{def}}{=}\mathcal{H}^{-1}\circ
\Delta\circ\mathcal{H}\text{,} \label{H(Delta)}%
\end{equation}
where
\[
\mathcal{H}=\iota^{\ast}\circ H^{\ast}\circ\rho
\]
Obviously, an analogous formula holds for the inverse map $H^{-1}$.
Furthermore, with any differential operator $\square:\mathcal{R}_{\mathbb{C}%
}\left(  J^{\infty}\left(  \pi\right)  \right)  \rightarrow\mathcal{R}%
_{\mathbb{C}}\left(  J^{\infty}\left(  \pi\right)  \right)  $ one associates
its real part $\square_{1}$ and its imaginary part $\square_{2}$, define by%
\[
\square_{j}\left(  f\right)  =\left(  \square\left(  f\right)  \right)
_{j}\text{,}%
\]
for any $f\in\mathcal{R}_{\mathbb{C}}\left(  J^{\infty}\left(  \pi\right)
\right)  $, $j=1,2$.

The linearization operators of equations (\ref{FirstEq}) and (\ref{SecondEq})
are related as follows:%
\begin{equation}
{\ell}_{\widetilde{F}}=H\left(  {\ell}_{F}\right)  \label{l_F_tilde}%
\end{equation}
Relation (\ref{l_F_tilde}) is an immediate consequence of $(\ref{cazzacchione}%
)$ and the identity $H\left(  D_{rx,sy}\right)  =\left(  H\left(
D_{x}\right)  \right)  ^{r}\circ\left(  H\left(  D_{y}\right)  \right)  ^{s}$.

\begin{remark}
\label{rem.che.mi.serve} In our case, the general scheme above applies to
equations (\ref{eqnostra}) and (\ref{eqShem}). More exactly, (\ref{eqShem}) is
the complex transformation of equation%
\begin{equation}
u_{XX}+u_{YY}+\frac{u_{X}}{X}=0\text{,} \label{eq_intermed}%
\end{equation}
which in its turn is obtained by equation (\ref{eqnostra}) by point
transformation
\begin{equation}
G:(x,y,u)\rightarrow(X,Y,u)\,,\quad\text{with}\quad X=x+y\,\,,\,\,Y=x-y.
\label{eq.trasformazione.intermedia}%
\end{equation}
\end{remark}

In the last part of this section we prove the main result of this paper.
Namely, we construct (see theorem \ref{zaza6}) a \textit{vector space}
isomorphism $\Theta$ from the algebra of real rational higher symmetries of
$\mathcal{E}$ into that of $\mathcal{Y}$.

%
%
%
%
%
%
%
%
%
%
%
%
%
%
%
%
%
%
%
%

\smallskip In order to prove theorem \ref{zaza6} some technical lemmata are necessary.

First of all, a straightforward computation shows that, for any vector field
$X$ on $J^{\infty}\left(  \pi\right)  $, the restrictions of $\left(
H^{-1}(X)\right)  _{1}$ and $\left(  H^{-1}(X)\right)  _{2}$ to $\mathcal{R}%
\left(  J^{\infty}\left(  \pi\right)  \right)  $ are vector fields on
$J^{\infty}(\pi)$.

\begin{lemma}
Let $\varphi\in\mathcal{R}\left(  J^{\infty}\left(  \pi\right)  \right)  $ be
the generating function of the evolutionary vector field $X_{\varphi}$. Then
\ we have that
\begin{equation}
H^{-1}(X_{\varphi})=X_{\mathcal{H}(\varphi)}\overset{\text{def}}{=}X_{\left(
\mathcal{H}(\varphi)\right)  _{1}}+iX_{\left(  \mathcal{H}(\varphi)\right)
_{2}} \label{zaza2}%
\end{equation}
\end{lemma}

\begin{proof}
Firstly $(H^{-1}(X_{\varphi}))_{1}$ and $(H^{-1}(X_{\varphi}))_{2}$ are
vertical vector fields. This is an immediate consequence of verticality of
$X_{\varphi}$ and equalities: $\mathcal{H}^{-1}\left(  x\right)  =\frac{1}%
{2}\left(  \xi+\eta\right)  $, $\mathcal{H}^{-1}\left(  y\right)  =\frac{i}%
{2}\left(  \eta-\xi\right)  $.

\smallskip Secondly, they are contact fields. In fact
\begin{align*}
\lbrack(H^{-1}(X_{\varphi}))_{j},D_{x}]  &  =[H^{-1}(X_{\varphi}),D_{x}%
]_{j}=\left(  H^{-1}[X_{\varphi},H(D_{x})]\right)  _{j}\\
&  =\left(  H^{-1}[X_{\varphi},D_{\xi}+D_{\eta}]\right)  _{j}\\
&  =\left(  H^{-1}[X_{\varphi},D_{\xi}]\right)  _{j}+\left(  H^{-1}%
[X_{\varphi},D_{\eta}]\right)  _{j}%
\end{align*}
and analogously%
\[
\lbrack(H^{-1}(X_{\varphi}))_{j},D_{y}]=\left(  H^{-1}[X_{\varphi},D_{\xi
}]\right)  _{j}-\left(  H^{-1}[X_{\varphi},D_{\eta}]\right)  _{j}.
\]

\smallskip Finally $(\mathcal{H}(\varphi))_{j}$ is the corresponding
generating section of $(H^{-1}(X_{\varphi}))_{j}$. In fact the generating
section of $(H^{-1}(X_{\varphi}))_{j}$ is $(H^{-1}(X_{\varphi}))_{j}\left(
u\right)  $, and
\begin{align*}
(H^{-1}(X_{\varphi}))_{j}(u)  &  =\left(  \mathcal{H}\circ X_{\varphi}%
\circ\mathcal{H}^{-1}\right)  _{j}(u)=(\mathcal{H}(X_{\varphi}(\mathcal{H}%
^{-1}(u))))_{j}\\
&  =(\mathcal{H}(X_{\varphi}(u)))_{j}=(\mathcal{H}(\varphi))_{j}\text{.}%
\end{align*}
\end{proof}

\smallskip Denote by $\operatorname*{Sym}_{\mathcal{R}}(\mathcal{Y})$ and
$\operatorname*{Sym}_{\mathcal{R}}(\mathcal{E})$, respectively, the algebra of
rational symmetries of $\mathcal{Y}$ and $\mathcal{E}$.

\begin{lemma}
\label{TheoSimmetrieComplesse}If $\varphi\in\operatorname*{Sym}_{\mathcal{R}%
}(\mathcal{Y})$ then $\left(  \mathcal{H}(\varphi)\right)  _{1}$ and $\left(
\mathcal{H}(\varphi)\right)  _{2}$ belong to $\operatorname*{Sym}%
_{\mathcal{R}}(\mathcal{E})$.
\end{lemma}

\begin{proof}
It is a direct consequence of (\ref{l_F_tilde}).
\end{proof}

\begin{lemma}
Let $\varphi\in\mathcal{R}(J^{\infty}(\pi))$. Then

\begin{itemize}
\item [1)]$\left(  \mathcal{H}^{-1}\left(  \mathcal{H}\left(  \varphi\right)
\right)  _{1}\right)  _{1}-\left(  \mathcal{H}^{-1}\left(  \mathcal{H}\left(
\varphi\right)  \right)  _{2}\right)  _{2}=\varphi$

\item[2)] $\left(  \mathcal{H}^{-1}\left(  \mathcal{H}\left(  \varphi\right)
\right)  _{1}\right)  _{2}=\left(  \mathcal{H}^{-1}\left(  \mathcal{H}\left(
\varphi\right)  \right)  _{2}\right)  _{1}=0$
\end{itemize}
\end{lemma}

\begin{proof}
We have that
\begin{align*}
\varphi &  =\mathcal{H}^{-1}\left(  \mathcal{H}\left(  \varphi\right)
\right)  =\mathcal{H}^{-1}\left(  \left(  \mathcal{H}\left(  \varphi\right)
\right)  _{1}+i\left(  \mathcal{H}\left(  \varphi\right)  \right)
_{2}\right)  =\mathcal{H}^{-1}\left(  \left(  \mathcal{H}\left(
\varphi\right)  \right)  _{1}\right)  +i\mathcal{H}^{-1}\left(  \left(
\mathcal{H}\left(  \varphi\right)  \right)  _{2}\right) \\
&  =\left(  \mathcal{H}^{-1}\left(  \mathcal{H}\left(  \varphi\right)
\right)  _{1}\right)  _{1}-\left(  \mathcal{H}^{-1}\left(  \mathcal{H}\left(
\varphi\right)  \right)  _{2}\right)  _{2}+i\left(  \left(  \mathcal{H}%
^{-1}\left(  \mathcal{H}\left(  \varphi\right)  \right)  _{1}\right)
_{2}+\left(  \mathcal{H}^{-1}\left(  \mathcal{H}\left(  \varphi\right)
\right)  _{2}\right)  _{1}\right)
\end{align*}
As $\varphi$ is a real function, relations 1) and
\begin{equation}
\left(  \mathcal{H}^{-1}\left(  \mathcal{H}\left(  \varphi\right)  \right)
_{1}\right)  _{2}+\left(  \mathcal{H}^{-1}\left(  \mathcal{H}\left(
\varphi\right)  \right)  _{2}\right)  _{1}=0 \label{eqaaaaaa}%
\end{equation}
hold. Also, $\mathcal{H}^{-1}\left(  \overline{\mathcal{H}\left(
\varphi\right)  }\right)  $ is real. In fact, first of all it holds:%
\begin{equation}
\overline{\mathcal{H}\left(  \varphi\right)  }=\overline{\mathcal{H}}\left(
\varphi\right)  \text{,} \label{zaza4}%
\end{equation}
where $\overline{\mathcal{H}}:\mathcal{R}_{\mathbb{C}}\left(  J^{\infty
}\left(  \pi\right)  \right)  \rightarrow\mathcal{R}_{\mathbb{C}}\left(
J^{\infty}\left(  \pi\right)  \right)  $ is the map associated with the
complex conjugate of matrix $H$. Then%
\[
\mathcal{H}^{-1}\left(  \overline{\mathcal{H}\left(  \varphi\right)  }\right)
=\mathcal{H}^{-1}\left(  \overline{\mathcal{H}}\left(  \varphi\right)
\right)  =\left(  \overline{H}\circ H^{-1}\right)  ^{\ast}\left(
\varphi\right)  \text{,}%
\]
which, keeping in mind that $\overline{H}\circ H^{-1}$ is a real
transformation (see section \ref{sec_remarkable_properties}), proves the
reality of $\mathcal{H}^{-1}\left(  \overline{\mathcal{H}\left(
\varphi\right)  }\right)  $. This implies that%
\[
\left(  \mathcal{H}^{-1}\left(  \mathcal{H}\left(  \varphi\right)  \right)
_{1}\right)  _{2}-\left(  \mathcal{H}^{-1}\left(  \mathcal{H}\left(
\varphi\right)  \right)  _{2}\right)  _{1}=0\text{.}%
\]
Then, in view of (\ref{eqaaaaaa}), we get
\[
\left(  \mathcal{H}^{-1}\left(  \mathcal{H}\left(  \varphi\right)  \right)
_{1}\right)  _{2}=\left(  \mathcal{H}^{-1}\left(  \mathcal{H}\left(
\varphi\right)  \right)  _{2}\right)  _{1}=0.\text{ \ \ \ \ \ }%
\]
\end{proof}

Now it is possible to prove the following

\begin{theorem}
\label{th.iniettiva} \label{zaza6}The linear map
\[
\Theta:\operatorname*{Sym}{}_{\mathcal{R}}(\mathcal{Y})\rightarrow
\operatorname*{Sym}{}_{\mathcal{R}}(\mathcal{E})\text{ , }\varphi
\mapsto\left(  \mathcal{H}\left(  \varphi\right)  \right)  _{1}+\left(
\mathcal{H}\left(  \varphi\right)  \right)  _{2}%
\]
is a vector space isomorphism. \ \ \
\end{theorem}

\begin{proof}
The map $\Theta^{\prime}:\eta\mapsto\left(  \mathcal{H}^{-1}\left(
\eta\right)  \right)  _{1}-\left(  \mathcal{H}^{-1}\left(  \eta\right)
\right)  _{2}$ is both the left and right inverse map of $\Theta$. In fact%
\begin{align*}
\Theta^{\prime}\left(  \Theta\left(  \varphi\right)  \right)   &  =\left(
\mathcal{H}^{-1}\left(  \mathcal{H}\left(  \varphi\right)  \right)
_{1}\right)  _{1}+\left(  \mathcal{H}^{-1}\left(  \mathcal{H}\left(
\varphi\right)  \right)  _{1}\right)  _{2}+\\
&  -\left(  \mathcal{H}^{-1}\left(  \mathcal{H}\left(  \varphi\right)
\right)  _{2}\right)  _{1}-\left(  \mathcal{H}^{-1}\left(  \mathcal{H}\left(
\varphi\right)  \right)  _{2}\right)  _{2}%
\end{align*}
which is equal to $\varphi$ in view of previous lemma. The same reasoning
holds for $\Theta\left(  \Theta^{\prime}\left(  \eta\right)  \right)  $.
\end{proof}

\begin{proposition}
Let $\varphi,\psi\in\mathcal{R}\left(  J^{\infty}(\pi)\right)  $. Then
$\mathcal{H}\left\{  \varphi,\psi\right\}  =\left\{  \mathcal{H}\left(
\varphi\right)  ,\mathcal{H}\left(  \psi\right)  \right\}  $.
\end{proposition}

\begin{proof}
The statement follows from definition (\ref{jacobrack}) and equation
(\ref{zaza2}).
\end{proof}

\begin{remark}
\label{rem.non.e.un.Lie.algebra.morfismo} The map $\Theta$ is not a Lie
algebra morphism, as a direct calculation shows.
\end{remark}

Now we shall reproduce similar results for recursion operators. Firstly,
denote by $\operatorname*{Rec}_{\mathcal{R}}(\mathcal{Y})$ and
$\operatorname*{Rec}_{\mathcal{R}}(\mathcal{E})$, respectively, the Lie
algebra of recursion operators of $\mathcal{Y}$ and $\mathcal{E}$ with
coefficients in $\mathcal{R}\left(  J^{\infty}(\pi)\right)  $. Similarly to
the case of symmetries, we have the following

\begin{proposition}
Let $\Re\in\operatorname*{Rec}_{\mathcal{R}}(\mathcal{Y})$. Then both $\left(
H^{-1}\left(  \Re\right)  \right)  _{1}$ and $\left(  H^{-1}\left(
\Re\right)  \right)  _{2}$ belong to $\operatorname*{Rec}_{\mathcal{R}%
}(\mathcal{E})$.
\end{proposition}

\begin{proof}
It is a straightforward application of definitions and of lemma
\ref{TheoSimmetrieComplesse}.
\end{proof}

\begin{theorem}
\label{isoRecur}The linear map%
\[
\Psi:\operatorname*{Rec}{}_{\mathcal{R}}(\mathcal{Y})\rightarrow
\operatorname*{Rec}{}_{\mathcal{R}}(\mathcal{E})\,\text{, }\Re\mapsto\left(
H^{-1}\left(  \Re\right)  \right)  _{1}+\left(  H^{-1}\left(  \Re\right)
\right)  _{2}%
\]
is a vector space isomorphism.
\end{theorem}

\begin{proof}
The proof is similar to the proof of theorem \ref{zaza6}, taking into account
that
\[
\left(  H\left(  H^{-1}\left(  \Re\right)  \right)  _{i}\right)  _{j}%
\varphi=\left(  \mathcal{H}^{-1}\left(  \mathcal{H}\left(  \Re\left(
\varphi\right)  \right)  \right)  _{i}\right)  _{j}\,,\quad i,j=1,2\text{.}%
\]
\end{proof}

\begin{remark}
\label{rem.point.invariance.of.principal.theorems} We note that all results
obtained in this section remain true if one replaces in (\ref{eq.H0}) matrix
$P$ (see (\ref{eq.P})) with the more general matrix $P\circ Q$ with $Q$ being
any $2\times2$ real matrix.
\end{remark}

\section{Examples of computations and Lie structure of rational symmetries of
$\mathcal{E}_{ED}$}

\label{sec.structure}

Below we apply the results obtained in the previous sections to $\mathcal{E}%
_{ED}$. In order to do this, in what follows by $H$ we mean the infinite
contact prolongation of $H_{0}\circ G$ where we recall that $H_{0}$ is defined
by (\ref{eq.H0}) and $G$ by (\ref{eq.trasformazione.intermedia}). The $H$ so
defined transforms $\mathcal{E}_{ED}^{\infty}$ into $\mathcal{Y}_{ED}^{\infty
}$ (see also remark \ref{rem.che.mi.serve}). Also, in view of remark
\ref{rem.point.invariance.of.principal.theorems}, all theorems of section
\ref{sec.transf.equ.and.its.symm} remain true for such $H$. We choose
$(x,y,u,u_{\xi},u_{\eta},\ldots,u_{k\xi},u_{k\eta},\ldots)$ as internal
coordinates on $\mathcal{Y}_{ED}^{\infty}$ and $(x,y,u,u_{x},u_{y}%
,\ldots,u_{x,(k-1)y},u_{ky},\ldots)$ as internal coordinates on $\mathcal{E}%
_{ED}^{\infty}.$

\smallskip As an example of computation of symmetries of $\mathcal{E}_{ED}$ we
give the following

\begin{proposition}
The symmetries of $\mathcal{E}_{ED}$ depending on derivatives up to second
order are linearly generated by the following ones:
\begin{align*}
X_{1} &  =(u_{x}+u_{y}+2u_{xy}x+2u_{xy}y)/(x+y)\\
X_{2} &  =(-yu_{x}+3xu_{y}-2u_{xy}y^{2}+2u_{xy}x^{2}+2u_{yy}x^{2}+4u_{yy}yx+\\
&  2u_{yy}y^{2}+xu_{x}+yu_{y})/(x+y)\\
X_{3} &  =(-u_{x}yx-u_{y}y^{2}+u_{y}x^{2}-u_{y}yx-2yx^{2}u_{xy}-2y^{2}%
xu_{xy}+u_{yy}x^{3}+\\
&  u_{yy}x^{2}y-u_{yy}xy^{2}-u_{yy}y^{3})/(x+y)\\
X_{4} &  =(3x^{2}yu_{x}+x^{3}u_{x}-3y^{3}u_{y}+uy^{2}-ux^{2}+9x^{2}%
yu_{y}-3xy^{2}u_{x}+8u_{xy}x^{3}y+\\
&  4u_{yy}x^{3}y+4u_{yy}y^{3}x-8u_{xy}y^{3}x+12u_{yy}x^{2}y^{2}+2u_{xy}%
x^{4}-2u_{xy}y^{4}-\\
&  2u_{yy}x^{4}-2u_{yy}y^{4}+3xy^{2}u_{y}-x^{3}u_{y}-y^{3}u_{x})/(x+y)\\
X_{5} &  =(uy^{3}+ux^{3}-12u_{xy}x^{2}y^{3}-12u_{xy}x^{3}y^{2}-20u_{y}%
y^{3}x+12u_{y}x^{3}y-\\
&  18u_{y}x^{2}y^{2}+4u_{x}y^{3}x-12u_{x}x^{3}y-18u_{x}x^{2}y^{2}%
-5uxy^{2}-5ux^{2}y+\\
&  u_{y}y^{4}+5u_{y}x^{4}+u_{x}x^{4}+5u_{x}y^{4}-8u_{yy}y^{4}x-8u_{yy}%
y^{3}x^{2}+8u_{yy}y^{2}x^{3}+\\
&  8u_{yy}yx^{4}+2u_{xy}y^{4}x+2u_{xy}x^{4}y+2u_{xy}x^{5}+2u_{xy}%
y^{5})/(x+y)\\
X_{6} &  =-u_{x}+u_{y}\\
X_{7} &  =u+2xu_{x}+2yu_{y}\\
X_{8} &  =ux-uy-u_{x}y^{2}+u_{x}x^{2}-2u_{x}yx+u_{y}x^{2}-u_{y}y^{2}%
+2u_{y}yx\\
X_{9} &  =u
\end{align*}
\end{proposition}

\begin{proof}
Firstly we take symmetries depending on second derivatives of $\mathcal{Y}%
_{ED}$, obtained in \cite{Shem94}, and then we transform them by using the map
$\Theta$ (see theorem \ref{th.iniettiva}).
\end{proof}

Symmetries $X_{6}$, $X_{7}$, $X_{8}$ and $X_{9}$ generate the algebra of
classical symmetries of $\mathcal{E}_{ED}$. We note that they are point
symmetries, as their generating sections are linear in the first derivatives.
This means that the corresponding vector fields on $J^{2}(\pi)$ are
prolongations of vector fields on $E$ rather than on $J^{1}(\pi)$. This fact
was noticed in \cite{CaMaPu05}, where variational aspects of $\mathcal{E}%
_{ED}$ were also studied.

Let us go back to equation $\mathcal{Y}_{ED}$. In \cite{Shem93},\cite{Shem94}%
,\cite{Shem95} it is proved that contact symmetries coincide with classical
point symmetries. More precisely we have that the most general generating
section of a contact symmetry is, up to an arbitrary solution, of the
following form:
\[
\varphi=\left(  c_{1}\frac{\left(  \eta-\xi\right)  }{2}+c_{4}\right)
u+\left(  -c_{1}\xi^{2}+c_{2}\xi-c_{3}\right)  u_{\xi}+\left(  c_{1}\eta
^{2}+c_{2}\eta+c_{3}\right)  u_{\eta}%
\]
where $c_{1},c_{2},c_{3},c_{4}$ are arbitrary constants. As equation
$\mathcal{Y}_{ED}$ is linear, such symmetries determine recursion operators as
they are linear in $u$ and in its derivatives (\cite{Olver01}). In particular
\begin{equation}
\square=D_{\xi}-D_{\eta},\quad\sigma=\xi D_{\xi}+\eta D_{\eta}+\frac
{\mathbb{I}}{2},\quad\tau=\xi^{2}D_{\xi}-\eta^{2}D_{\eta}+\frac{\xi-\eta}%
{2}\mathbb{I} \label{RecursionOperators}%
\end{equation}
are recursion operators. Since higher symmetries of $\mathcal{Y}_{ED}$ are
linear in $u$ and in its derivatives (see theorem
\ref{th.linearita.Shemarulin} below), then in this case the theory of higher
symmetries can be developed using recursion operators as fundamental objects
(\cite{Olver01}). For instance, it is natural to ask if by applying arbitrary
compositions of (\ref{RecursionOperators}) to the symmetry $u$ we get the
whole algebra of higher symmetries. For this purpose, let us consider
\begin{equation}
\square_{j}^{m}=[\ldots\lbrack\square^{m},\underset{\text{j-times}%
}{\underbrace{\tau],\ldots,\tau}}] \label{RecursionOperators2}%
\end{equation}
and
\[
\varphi_{0}=\square(u),\quad\varphi_{1}=\tau(u),\quad\varphi_{2}=\xi u_{\xi
\xi}-\eta u_{\eta\eta}+\frac{\xi u_{\xi}-\eta u_{\eta}}{\xi+\eta}\text{.}%
\]
Since the recursion operators (\ref{RecursionOperators}) are $\mathcal{C}%
$-differential operators, it is well defined the restriction $\overline
{\square}_{j}^{m}$ of (\ref{RecursionOperators2}) on the equation. The
following three theorems are due to Shemarulin (see \cite{Shem93}%
,\cite{Shem94},\cite{Shem95}).

\begin{theorem}
\label{th.linearita.Shemarulin} Let $\varphi\in C^{\infty}\left(
\mathcal{Y}_{ED}^{n-2}\right)  $ be a symmetry of $\mathcal{Y}_{ED}$. Then
$\varphi$ has the following form:
\begin{equation}
\varphi=\phi(\xi,\eta)+\sum_{0\leq k\leq n}\mathcal{P}_{k}(\xi,\eta)u_{k\xi
}+\sum_{0\leq h\leq n}\mathcal{Q}_{h}(\xi,\eta)u_{h\eta}%
,\label{espressione_lineare_phi_nostra}%
\end{equation}
where $\phi$ is a solution of $\mathcal{Y}_{ED}$ and $\mathcal{P}_{k}$ and
$\mathcal{Q}_{h}$ are rational functions.
\end{theorem}

\begin{theorem}
\label{theoFinitamenteGenerata} The algebra $\operatorname*{Sym}\left(
\mathcal{Y}_{ED}\right)  $ is the semi-direct sum $A\oplus\operatorname*{NSym}%
\left(  \mathcal{Y}_{ED}\right)  $ where $A$ is the abelian infinite
dimensional ideal of solutions of $\mathcal{Y}_{ED}$ and $\operatorname*{NSym}%
\left(  \mathcal{Y}_{ED}\right)  $ is the algebra linearly generated by $u$
and $\varphi_{j}^{m}$ where
\[
\varphi_{j}^{m}=\{\ldots\{\{\ldots\{\varphi_{0},\underset{\text{(j-1)-times}%
}{\underbrace{\varphi_{2}\}\ldots\varphi_{2}}}\}\underset{\text{m-times}%
}{\underbrace{\varphi_{1}\}\ldots\varphi_{1}}}\}\text{.}%
\]
\end{theorem}

\begin{theorem}\label{TheoShemCommutatori}
The algebra $\operatorname*{NSym}\left(  \mathcal{Y}_{ED}\right)
$ is linearly generated by $u$ and
$\overline{\square}_{j}^{m}(u)$. Moreover, we
have the following relations:%
\begin{align*}
\left[  \square_{j}^{m},\square\right]   &  =-j\left(  2m-j+1\right)
\square_{j-1}^{m}, & 1 &  \leq j\leq2m;\\
\left[  \square_{j}^{m},\sigma\right]   &  =\left(  m-j\right)  \square
_{j}^{m}, & 0 &  \leq j\leq2m;\\
\left[  \square_{j}^{m},\tau\right]   &  =\square_{j+1}^{m}, & 0 &  \leq
j\leq2m;
\end{align*}
and
\begin{align*}
\overline{\square}_{j}^{m}(u) &  \neq0, & 0 &  \leq j\leq2m;\\
\overline{\square}_{j}^{m}(u) &  =0, & j &  \geq2m+1.
\end{align*}
\end{theorem}

Now we shall reproduce, for the equation $\mathcal{E}_{ED}$, similar results
by using the complex transformation $H$.

\begin{proposition}
\label{prop.linearita.delle.nostre.simmetrie} Rational symmetries of
$\mathcal{E}_{ED}$ are linear in the internal jet variables.
\end{proposition}

\begin{proof}
We recall (theorem \ref{zaza6}) that rational symmetries of
$\mathcal{E}_{ED}$ are images through $\mathcal{H}$ of symmetries
of $\mathcal{Y}_{ED}$, which have the form
(\ref{espressione_lineare_phi_nostra}). In view of the fact $H$ is
a block matrix, the proposition follows if we show that the
restriction of $u_{mx,ny}$ to $\mathcal{E}_{ED}^{m+n-2}$ is a
linear function in the internal jet variables. We show it by
induction. Firstly, a straightforward computation shows that
\begin{equation}
\left.  u_{xx,hy}\right|  _{\mathcal{E}_{ED}^{h}}=-\sum_{k=0}^{h}\binom{h}%
{k}\frac{(-1)^{k}k!}{(x+y)^{k+1}}\left(  u_{x,(h-k)y}+u_{(h-k+1)y}\right)
-u_{(h+2)y}.\label{eq.base.di.induzione}%
\end{equation}
Now let us suppose that $\left.  u_{(m-1)x,ny}\right|  _{\mathcal{E}%
_{ED}^{m+n-3}}$ is linear in the internal jet variables. Namely%
\[
\left.  u_{(m-1)x,ny}\right|  _{\mathcal{E}_{ED}^{m+n-3}}=\sum_{h=0}%
^{m+n-2}a_{h}(x,y)u_{x,hy}+\sum_{j=0}^{m+n-1}b_{j}(x,y)u_{jy}.
\]
Then%
\begin{align*}
\left.  u_{mx,ny}\right|  _{\mathcal{E}_{ED}^{m+n-2}} &  =\bar{D}_{x}\left(
\left.  u_{(m-1)x,ny}\right|  _{\mathcal{E}_{ED}^{m+n-1}}\right)  \\
&  =\sum_{h=0}^{m+n-2}\left(  \frac{\partial a_{h}}{\partial x}u_{x,hy}%
+a_{h}\left.  u_{xx,hy}\right|  _{\mathcal{E}_{ED}^{h}}\right)  +\sum
_{j=0}^{m+n-1}\left(  \frac{\partial b_{j}}{\partial x}u_{jy}+b_{j}%
u_{x,jy}\right) .
\end{align*}
The assertion is proved in view of (\ref{eq.base.di.induzione}).
\end{proof}

\smallskip We define the algebra $\operatorname*{NSym}\left(  \mathcal{E}%
_{ED}\right)  $ as the algebra of rational higher symmetries of $\mathcal{E}%
_{ED}$ up to solutions of $\mathcal{E}_{ED}$. As we have noticed in remark
\ref{rem.non.e.un.Lie.algebra.morfismo}, the map $\Theta$ is not a Lie algebra
morphism. Anyway we have the following

\begin{proposition}
The algebra $\operatorname*{NSym}\left(  \mathcal{E}_{ED}\right)  $ is
infinite dimensional as vector space and finitely generated as Lie algebra.
More precisely
\[
\varrho_{j}^{m}=\{\ldots\{\{\ldots\{\varrho_{0},\underset{\text{(j-1)-times}%
}{\underbrace{\varrho_{2}\}\ldots\varrho_{2}}}\}\underset{\text{m-times}%
}{\underbrace{\varrho_{1}\}\ldots\varrho_{1}}}\}
\]
where
\begin{align*}
\varrho_{0} &  =-u_{x}+u_{y}\\
\varrho_{1} &  =\frac{1}{2}\left(  x^{2}-2xy-y^{2}\right)  u_{x}+\frac{1}%
{2}\left(  x^{2}+2xy-y^{2}\right)  u_{y}+\frac{1}{2}\left(  x-y\right)  u\\
\varrho_{2} &  =\left(  x+y\right)  u_{yy}+\left(  x-y\right)  u_{xy}+\frac
{1}{2\left(  x+y\right)  }\left(  \left(  x-y\right)  u_{x}+\left(
3x+y\right)  u_{y}\right)
\end{align*}
form, together with $u$, a linear basis of $\operatorname*{NSym}\left(
\mathcal{E}_{ED}\right)  $.
\end{proposition}

\begin{proof}
The proposition follows taking into consideration that $\Theta$ restricts on
$\operatorname*{NSym}\left(  \mathcal{Y}_{ED}\right)  $ and
$\operatorname*{NSym}\left(  \mathcal{E}_{ED}\right)  $, theorem
\ref{theoFinitamenteGenerata}, and finally that $\mathcal{H}\left(
\varphi_{0}\right)  =i\varrho_{0}$, $\mathcal{H}\left(  \varphi_{1}\right)
=i\varrho_{1}$, $\mathcal{H}\left(  \varphi_{2}\right)  =i\varrho_{2}$.
\end{proof}

\medskip By transforming recursion operators (\ref{RecursionOperators})
throughout $H^{-1}$, we get the following

\begin{proposition}%
\[
H^{-1}(\square)=i\widetilde{\square},\quad H^{-1}(\tau)=i\widetilde{\tau
},\quad H^{-1}(\sigma)=\widetilde{\sigma},
\]
where%
\begin{align*}
\widetilde{\square}  &  =-D_{x}+D_{y}\\
\widetilde{\tau}  &  =\left(  \frac{1}{2}x^{2}-xy-\frac{1}{2}y^{2}\right)
D_{x}+\left(  \frac{1}{2}x^{2}+xy-\frac{1}{2}y^{2}\right)  D_{y}+\frac{1}%
{2}\left(  x-y\right)  \mathbb{I}\\
\widetilde{\sigma}  &  =xD_{x}+yD_{y}+\frac{\mathbb{I}}{2}%
\end{align*}
\end{proposition}

Now, if we define
\[
\bigtriangledown_{j}^{m}=[\ldots\lbrack\widetilde{\square}^{m},\underset
{\text{j-times}}{\underbrace{\widetilde{\tau}],\ldots,\widetilde{\tau}}}]
\]
we get the Lie structure of $\operatorname*{NSym}\left(  \mathcal{E}%
_{ED}\right)  $ by means of the following

\begin{theorem}
The algebra $\operatorname*{NSym}\left(  \mathcal{E}_{ED}\right)  $ is
linearly generated by $u$ and $\overline{\nabla}_{j}^{m}(u)$. Moreover, we
have the following relations:
\begin{align*}
\left[  \nabla_{j}^{m},\widetilde{\square}\right]   &  =j\left(
2m-j+1\right)  \nabla_{j-1}^{m}, & 1  &  \leq j\leq2m;\\
\left[  \nabla_{j}^{m},\widetilde{\sigma}\right]   &  =\left(  m-j\right)
\nabla_{j}^{m}, & 0  &  \leq j\leq2m;\\
\left[  \nabla_{j}^{m},\widetilde{\tau}\right]   &  =\nabla_{j+1}^{m}, & 0  &
\leq j\leq2m;
\end{align*}
and
\begin{align*}
\overline{\nabla}_{j}^{m}(u)  &  \neq0, & 0  &  \leq j\leq2m;\\
\overline{\nabla}_{j}^{m}(u)  &  =0, & j  &  \geq2m+1.
\end{align*}
\end{theorem}

\begin{proof}
We have that
\begin{align*}
H^{-1}\left(  \square_{j}^{m}\right)   &  =[\ldots\lbrack\left(  H^{-1}\left(
\square\right)  \right)  ^{m},\underset{\text{j-times}}{\underbrace
{H^{-1}\left(  \tau\right)  ],\ldots,H^{-1}\left(  \tau\right)  }}]\\
&  =i^{m+j}[\ldots\lbrack\widetilde{\square}^{m},\underset{\text{j-times}%
}{\underbrace{\widetilde{\tau}],\ldots,\widetilde{\tau}}}]=i^{m+j}%
\bigtriangledown_{j}^{m}\text{.}%
\end{align*}
Then, taking into account theorem \ref{isoRecur} and theorem
\ref{TheoShemCommutatori}, the theorem follows.
\end{proof}

\medskip Finally, taking into account that symmetries of $\mathcal{E}_{ED}$
are linear in $u_{\sigma}$ (see proposition
\ref{prop.linearita.delle.nostre.simmetrie}), for each couple $\Delta,\nabla$
of recursion operators of $\mathcal{E}_{ED}$, we have that $[\Delta
,\nabla](u)=-\{\Delta(u),\nabla(u)\}$. Then, in view of previous theorem, we
get the Lie structure of $\operatorname*{NSym}\left(  \mathcal{E}_{ED}\right)
$.

\bigskip

\textbf{Acknowledgements.} We wish to thank A.M. Vinogradov for having
proposed the problem. We also thank R. Alonso Blanco, E. Barletta, S. Dragomir
and M. Marvan for useful suggestions and stimulating discussions. Finally we
wish to thank warmly V. Rossi for his constant support.

\end{document}